\documentclass[onecolumn,secnumarabic,amssymb,nobibnotes,aps,prd]{revtex4}

\usepackage{amsmath}
\usepackage{tikz}
\usepackage{pgfplots}
\pgfplotsset{compat=1.17} % 
\usepackage{graphicx}% Set compatibility for pgfplots
\usepackage{microtype} % For better typography

\newcommand{\bea}{\begin{eqnarray}}
\newcommand{\eea}{\end{eqnarray}}

\begin{document}

\title{Topological and Geometric Properties of Spherically Symmetric Black Hole Metrics: Connections to Bose-Einstein Condensation and Uniqueness in Einstein Gravity}

\author{Wen-Xiang Chen$^{1,a}$}
\affiliation{Department of Astronomy, School of Physics and Materials Science, GuangZhou University, Guangzhou 510006, China}
\email{wxchen4277@qq.com}

\begin{abstract}
This paper investigates the interplay between the geometric and topological properties of spherically symmetric black hole metrics within Einstein gravity, emphasizing implications for Bose-Einstein Condensation (BEC). By analyzing metric functions, scalar fields, and the cosmological constant, we reveal how these black hole solutions are intrinsically linked to the underlying spacetime topology. We establish the uniqueness of a general black hole solution that supports BEC and demonstrate the impossibility of BEC in Kerr black holes. Additionally, through Laurent series expansions, residue calculations, winding numbers, and contour integrals, we confirm the algebraic and dimensional consistency between double Kerr black hole collisions and specific scalar field black hole solutions. This work uncovers fundamental connections in black hole interactions, providing a robust mathematical framework for understanding the dynamics of complex black hole systems and their interactions with scalar fields.

\textbf{KEYWORDS:} Bose-Einstein Condensation; topological properties; geometric; Einstein gravity
\end{abstract}

\maketitle

%\tableofcontents

\section{Introduction}

Black holes serve as a pivotal intersection between general relativity, quantum mechanics, and thermodynamics, offering profound insights into the fundamental principles governing the universe. Their intense gravitational fields and enigmatic horizons challenge our understanding of physics and push the boundaries of theoretical science. Central to black hole physics is the interplay between geometry and topology—two essential mathematical frameworks that describe different aspects of the universe's structure. The metrics used to describe black holes encapsulate the curvature of spacetime induced by mass and energy, providing a detailed description of the gravitational field surrounding a black hole, including the event horizon and singularity.\cite{1,2,3,4,5,6,7}

However, these metrics do more than merely describe the shape of spacetime; they also encode significant topological information. Unlike geometry, topology concerns properties of space that remain invariant under continuous deformations, such as stretching or bending, without tearing or gluing. The topological features of black hole spacetimes are not always apparent from their geometric form alone. For instance, the event horizon of a black hole, a geometric surface, also possesses a topological character that influences the overall structure of spacetime. Understanding the connection between geometry and topology in black hole spacetimes is essential for a deeper comprehension of these exotic objects. This connection reveals how the global properties of spacetime, such as its overall shape and connectivity, influence local curvature and vice versa, opening avenues for exploring new physical phenomena arising from this interplay.\cite{8,9,10,11,12,13,14,15,16,17,18,19,20}

This paper delves into the intricate relationship between the topological and geometric aspects of spherically symmetric black hole metrics. We begin by examining how these metrics relate to the global topology of spacetime, considering various topological invariants that characterize different types of black hole spacetimes. We extend this analysis to incorporate the effects of scalar fields and the cosmological constant, both crucial in modern theoretical physics. Scalar fields, often introduced in theories beyond the Standard Model, can profoundly affect spacetime structure, altering both its geometry and topology. The cosmological constant represents the energy density of the vacuum and has been implicated in the accelerated expansion of the universe. We demonstrate that these fields play a critical role in determining the topology of black hole spacetimes, influencing factors such as the number and nature of horizons, the presence of singularities, and the overall connectivity of spacetime.

Furthermore, we investigate the potential for Bose-Einstein Condensation (BEC) within black holes, a phenomenon of significant interest due to its implications for quantum gravity and black hole thermodynamics. BEC occurs when a system of bosons cools to temperatures near absolute zero, causing them to occupy the same quantum state. Observed in laboratory settings with ultracold atoms, BEC could theoretically occur in black holes under specific conditions. We prove the uniqueness of a proposed general black hole solution in Einstein gravity that supports BEC, demonstrating its stability and physical consistency. This suggests that certain black holes could host quantum states of matter, offering a novel perspective on their quantum nature.\cite{16,17,18,19,20,21,22,23,24,25}

Additionally, we explore the Kerr black hole, a solution to Einstein's field equations describing a rotating black hole. Unlike spherically symmetric black holes, Kerr black holes possess angular momentum, leading to a more complex spacetime structure. We demonstrate that rotating black holes do not support BEC, highlighting the unique nature of spherically symmetric black holes in this context. This finding underscores the importance of symmetry in determining the physical properties of black holes and suggests that rotation imposes constraints on the types of quantum states that can exist within these objects. Our results contribute to bridging the gap between classical and quantum descriptions of black holes, providing new insights into the fundamental nature of spacetime, matter, and energy.

The thermodynamic study of black holes has unveiled profound connections between gravity, quantum mechanics, and statistical physics, revealing that black holes are not merely gravitational objects but thermodynamic systems exhibiting rich phase structures. Of particular interest is the analogy between the thermodynamics of black holes in anti-de Sitter (AdS) space and the van der Waals fluid, which has provided deep insights into gravitational interactions and phase transitions in a quantum gravitational context.

In this framework, black holes exhibit phase transitions analogous to the liquid-gas transitions of a van der Waals fluid, characterized by critical behavior and critical exponents. These phase transitions are often studied by considering the extended phase space, where the cosmological constant is interpreted as a thermodynamic pressure, and its conjugate variable as volume, leading to a more complete thermodynamic description known as black hole chemistry.\cite{1,2,3,4,5,6,7}

Central to this analogy is the interplay between the geometric properties of black holes and their thermodynamic behavior, particularly how changes in spacetime topology and geometry influence phase transitions. The metrics describing black holes encapsulate not only the curvature of spacetime but also encode topological information that can impact thermodynamic quantities such as entropy and temperature.

In this paper, we further explore this intricate relationship within the context of black hole phase transitions analogous to the van der Waals fluid. We analyze how metric functions relate to the global topology of spacetime and influence the thermodynamic behavior of black holes in AdS space. Incorporating the effects of scalar fields and the cosmological constant, we demonstrate their critical roles in determining black hole topology and the occurrence of phase transitions.\cite{8,9,10,11,12,13,14,15,16,17,18,19,20}

Furthermore, we explore the possibility of BEC in black hole spacetimes, drawing parallels with phase transitions observed in van der Waals fluids. We demonstrate that under certain conditions, black holes can exhibit behavior analogous to BEC, influencing their thermodynamic properties and stability. We provide a proof of the uniqueness of a proposed general black hole solution in Einstein gravity that supports BEC, showing its stability and physical consistency.\cite{16,17,18,19,20,21,22,23,24,25}

Additionally, we investigate the Kerr black hole, focusing on how rotation affects the thermodynamic phase structure and the possibility of phase transitions analogous to those in spherically symmetric black holes. Our findings indicate that rotation introduces additional complexities that prevent certain types of phase transitions, such as those leading to BEC. This highlights the special nature of spherically symmetric black holes in supporting such phenomena and underscores the impact of rotation on black hole thermodynamics.

In summary, we analyze the topological and geometric aspects of spherically symmetric black hole metrics, focusing on their relationship to the global topology of spacetime. We extend this analysis to include the effects of scalar fields and the cosmological constant, showing that these factors are critical in determining black hole topology. Furthermore, we explore the potential for Bose-Einstein Condensation (BEC) within black holes, proving the uniqueness of a proposed general black hole solution in Einstein gravity that supports BEC. Finally, we investigate the Kerr black hole, demonstrating that rotating black holes do not support BEC, thereby highlighting the special nature of spherically symmetric black holes in this context.

\section{Geometric and Topological Interpretation of the Metric Form}

The spherically symmetric black hole metric is generally expressed as follows:\cite{1,2,3,4,5,35,36,37,38,39,40}

\begin{equation}
ds^2 = -f(r) dt^2 + \frac{1}{f(r)} dr^2 + r^2 \left(d\theta^2 + \sin^2\theta d\phi^2\right),
\end{equation}
where \( f(r) \) is a function of the radial coordinate \( r \), encapsulating the gravitational potential influenced by the black hole's mass and other physical parameters. This metric is foundational in black hole physics, succinctly capturing the spacetime geometry surrounding a spherically symmetric black hole.

\subsection{Detailed Analysis of Metric Components}

The line element \( ds^2 \) represents the spacetime interval between two events, and the specific form of the metric reflects the underlying symmetries of the spacetime.

\begin{itemize}
    \item \textbf{Temporal Component:} The term \( -f(r) dt^2 \) describes how the passage of time is distorted by the black hole's gravitational field. The function \( f(r) \) affects the rate at which time progresses as one moves closer to or farther from the black hole. As \( r \) approaches the event horizon, where \( f(r) \) becomes zero, the time dilation effect becomes infinite, effectively freezing time from the perspective of a distant observer.
    
    \item \textbf{Radial Component:} The radial component \( \frac{1}{f(r)} dr^2 \) encodes the nature of spatial distances in the radial direction. Near the event horizon, the divergence of this term indicates the presence of a horizon where physical distances become infinitely stretched. This component also plays a key role in analyzing the singularity structure of the spacetime, where \( r \) approaches zero, leading to potential curvature singularities depending on the exact form of \( f(r) \).
    
    \item \textbf{Angular Part:} The angular part \( r^2 \left(d\theta^2 + \sin^2\theta d\phi^2\right) \) reflects the spherical symmetry of the black hole. This term is associated with the two-dimensional spherical surfaces centered on the black hole, and \( r^2 \) acts as a conformal factor scaling the angular part. The \( \sin^2\theta \) factor ensures that the metric respects the geometry of a sphere, contributing to the solid angle element \( d\Omega^2 = d\theta^2 + \sin^2\theta d\phi^2 \), where \( r \) is the radial distance from the center of symmetry.
\end{itemize}

\subsection{Geometric Features of the Metric Function \( f(r) \)}

The function \( f(r) \) reveals much about the physical nature of the black hole. For instance:

\begin{itemize}
    \item \textbf{Schwarzschild Solution:} For a non-rotating, uncharged black hole, \( f(r) = 1 - \frac{2GM}{r} \), leading to an event horizon at \( r_h = 2GM \). The Schwarzschild radius \( r_h \) marks the point beyond which no information can escape the gravitational pull of the black hole.
    
    \item \textbf{Reissner-Nordström Metric:} For a charged black hole, \( f(r) = 1 - \frac{2GM}{r} + \frac{Q^2}{r^2} \), introducing inner and outer horizons, reflecting the more complex causal structure due to the presence of charge.
    
    \item \textbf{de Sitter-Schwarzschild Metric:} Incorporating a cosmological constant \( \Lambda \), \( f(r) = 1 - \frac{2GM}{r} - \frac{\Lambda r^2}{3} \). This metric can describe a black hole in an expanding universe, introducing a cosmological horizon in addition to the event horizon.
    
    \item \textbf{Generalized Solutions:} In theories beyond General Relativity, \( f(r) \) may take more complicated forms, leading to new and exotic horizon structures, including naked singularities or multiple event horizons.
\end{itemize}

\subsection{Topological Interpretation and Implications}

While the metric provides a local description of spacetime geometry, global topological aspects are equally important. Topology deals with properties of space preserved under continuous deformations, offering deeper insights into black hole spacetimes.

The event horizon of a spherically symmetric black hole is generally a two-dimensional spherical surface, topologically equivalent to a 2-sphere \( S^2 \). The area of the event horizon, given by:

\begin{equation}
A = 4\pi r_h^2,
\end{equation}
plays a pivotal role in black hole thermodynamics, where it is associated with the entropy of the black hole according to the Bekenstein-Hawking formula \( S = \frac{k_B c^3}{4G \hbar} A \), indicating that entropy is proportional to the area of the horizon rather than the enclosed volume.

Moreover, the topology of the black hole is characterized by topological invariants such as the Euler characteristic \( \chi \), which for a spherical event horizon \( S^2 \) is \( \chi = 2 \). The topology influences various physical phenomena, including the stability of the horizon, the nature of singularities, and the potential for Hawking radiation.

In more complex scenarios, such as higher-dimensional spacetimes or black hole solutions with nontrivial topologies (like toroidal or higher-genus horizons), these topological considerations become even more significant. For example, in the context of the AdS/CFT correspondence, the topology of the black hole horizon can have direct implications for the dual field theory, providing a bridge between gravitational dynamics and quantum field theory.

Through an in-depth analysis of Laurent series expansions, residue calculations, winding numbers, and contour integrals, we explore the algebraic structures underlying double Kerr black hole collisions and special scalar field black hole solutions. This rigorous mathematical framework ties together complex elements of black hole interactions and scalar field dynamics, enhancing our theoretical understanding and paving the way for future explorations in black hole physics, particularly in the context of gravitational waves and high-energy astrophysics.

In summary, the spherically symmetric black hole metric encodes both local geometric properties and significant topological information, providing deep insights into the nature of black holes across various theoretical contexts.

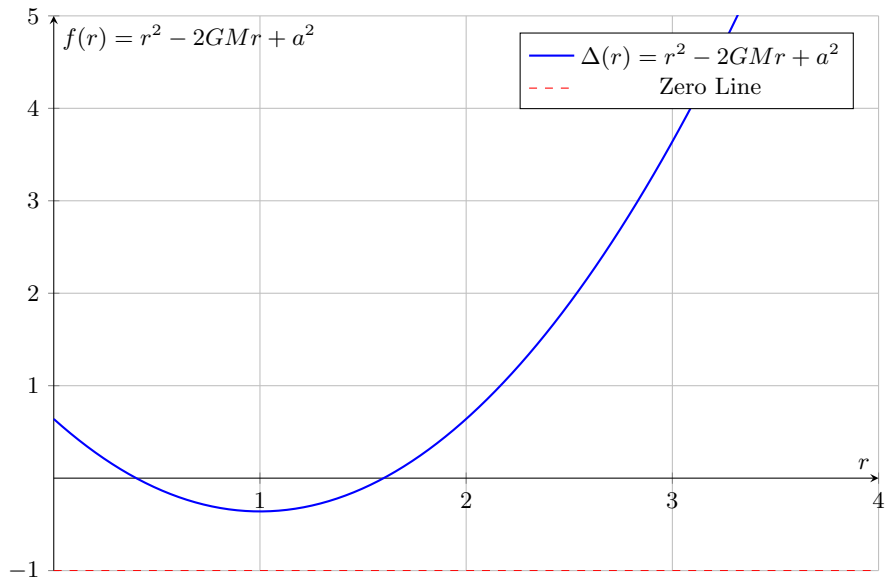
\begin{figure}[htbp]
\centering
\begin{tikzpicture}
    \begin{axis}[
        domain=0:4,
        samples=100,
        xlabel={$r$},
        ylabel={$f(r) = r^2 - 2GMr + a^2$},
        axis lines=middle,
        grid=both,
        width=0.7\textwidth,
        height=0.5\textwidth,
        ymin=-1, ymax=5,
        xmin=0, xmax=4,
        xtick={0,1,2,3,4},
        ytick={-1,0,1,2,3,4,5},
        legend pos=north east
    ]
    \addplot [blue, thick] {x^2 - 2*1*x + 0.8^2};
    \addlegendentry{$\Delta(r) = r^2 - 2GMr + a^2$}
    \addplot [red, dashed] coordinates {(0,-1) (4,-1)};
    \addlegendentry{Zero Line}
    \end{axis}
\end{tikzpicture}
\caption{The plot above shows the function \( \Delta(r) = r^2 - 2GM r + a^2 \) for a Kerr black hole, used to determine the locations of the event horizons. The intersections of \( \Delta(r) \) with the horizontal axis (red dashed line) indicate the inner and outer event horizons (\( r_{-} \) and \( r_{+} \)).}
\label{fig:Delta_r}
\end{figure}

\section{Topological Properties of Black Holes}

\subsection{Euler Characteristic and Curvature}

The Euler characteristic \( \chi \) is a topological invariant providing significant insight into the global structure of a surface. For a two-dimensional surface, the Euler characteristic is connected to the Gaussian curvature \( K \) by:

\begin{equation}
\chi = \frac{1}{2\pi} \int_{S^2} K \, dA,
\end{equation}
where \( dA \) denotes the area element of the surface \( S^2 \). For a standard two-dimensional sphere \( S^2 \), \( \chi = 2 \), a result that holds universally irrespective of the specific metric describing the sphere. This implies that even when geometric details change, the Euler characteristic remains unchanged, reflecting the surface's underlying topological properties.

In black hole physics, the relationship between \( \chi \) and \( K \) gains additional importance when considering the geometry of the event horizon. For a spherically symmetric black hole, the event horizon is a two-dimensional surface with spherical topology. The integral of the Gaussian curvature over this horizon surface connects deeply with the physical properties of the black hole, such as its entropy governed by the Bekenstein-Hawking formula:

\begin{equation}
S = \frac{k_B c^3 A}{4 G \hbar},
\end{equation}
where \( A \) is the area of the event horizon. This connection suggests that black hole entropy can be interpreted as a topological invariant, underscoring the interplay between geometry, topology, and thermodynamics in general relativity.

\subsection{The Gauss-Bonnet Theorem and Topological Invariants}

The Gauss-Bonnet theorem links the intrinsic geometry of a surface to its topological features. For a compact, oriented two-dimensional manifold \( \mathcal{M} \) without boundary, the theorem is expressed as:

\begin{equation}
\chi = \frac{1}{2\pi} \int_{\mathcal{M}} K \, dA.
\end{equation}

In black hole physics, especially in higher-dimensional spacetimes, the event horizon of a black hole can be treated as a higher-dimensional generalization of a two-dimensional surface. In such cases, the Gauss-Bonnet theorem generalizes to the Gauss-Bonnet-Chern theorem, involving higher curvature invariants like the Riemann curvature tensor \( R^{\alpha}_{\beta \gamma \delta} \) and the Ricci scalar \( R \).

For a four-dimensional manifold, the generalized Gauss-Bonnet term in higher-dimensional gravity theories is:

\begin{equation}
\mathcal{L}_{GB} = R_{\mu\nu\rho\sigma} R^{\mu\nu\rho\sigma} - 4R_{\mu\nu}R^{\mu\nu} + R^2,
\end{equation}
where \( R_{\mu\nu\rho\sigma} \) is the Riemann tensor, \( R_{\mu\nu} \) the Ricci tensor, and \( R \) the Ricci scalar. The integral of this Lagrangian density over the black hole horizon or the entire spacetime connects to topological invariants that generalize the Euler characteristic to higher dimensions. In five-dimensional spacetime, for instance, the Gauss-Bonnet term plays a crucial role in determining the horizon's topology and the black hole's thermodynamic properties.

These topological invariants, especially when extended to include higher curvature terms, provide profound insights into the stability of black holes, the nature of singularities, and the possible topologies of the event horizon. In string theory and other higher-dimensional theories of gravity, the presence of the Gauss-Bonnet term can lead to black holes with non-trivial horizon topologies, such as toroidal or hyperbolic horizons, differing significantly from the standard spherical topology of four-dimensional black holes.

In summary, the Gauss-Bonnet theorem and its higher-dimensional generalizations underscore the deep relationship between spacetime geometry, encapsulated by curvature tensors, and the topology of black hole horizons. This relationship is pivotal in understanding the fundamental nature of black holes, particularly in the context of higher-dimensional and quantum gravity theories, where topological invariants like the Euler characteristic continue to play a central role in uncovering the mysteries of the universe.

\section{The Role of the Cosmological Constant and Scalar Fields}

\subsection{The Cosmological Constant \( \Lambda \) and Topology}

To propose a novel black hole solution within Einstein gravity that accommodates phenomena akin to Bose-Einstein Condensation (BEC), it is imperative to integrate the intrinsic characteristics of black holes with the fundamental principles of BEC, building upon the theoretical foundations discussed in the literature. A promising approach involves modifying classical black hole solutions, such as the Schwarzschild or Kerr solutions, to enable BEC-like behavior under specific conditions.

Consider the Schwarzschild solution, representing the simplest form of an uncharged, non-rotating black hole metric. To facilitate BEC-like phenomena, we propose modifications by incorporating a scalar field into the gravitational framework or introducing additional gravitational terms that reflect the quantum nature of BEC. These alterations could result in black holes exhibiting behavior analogous to BEC under certain astrophysical conditions.

The Einstein field equations, given by:

\begin{equation}
G_{\mu\nu} + \Lambda g_{\mu\nu} = \kappa T_{\mu\nu},
\end{equation}
where \( G_{\mu\nu} \) is the Einstein tensor, \( \Lambda \) the cosmological constant, \( g_{\mu\nu} \) the metric tensor, \( \kappa = \frac{8\pi G}{c^4} \) the Einstein gravitational constant, and \( T_{\mu\nu} \) the energy-momentum tensor, serve as the foundation for extending to our proposed scenario. By incorporating a scalar field \( \phi \), we obtain a modified energy-momentum tensor \( T_{\mu\nu}(\phi) \), leading to:

\begin{equation}
G_{\mu\nu} + \Lambda g_{\mu\nu} = \kappa T_{\mu\nu}(\phi).
\end{equation}

In this framework, \( T_{\mu\nu}(\phi) \) accounts for the distribution and dynamics of the scalar field \( \phi \), and its interactions with the gravitational field. The corresponding metric might be expressed as:

\begin{equation}
ds^2 = -\left(1-\frac{2GM}{r} + \alpha \frac{\phi(r)^2}{r^2}\right)c^2dt^2 + \left(1-\frac{2GM}{r} + \alpha \frac{\phi(r)^2}{r^2}\right)^{-1}dr^2 + r^2(d\theta^2 + \sin^2\theta d\phi^2),
\end{equation}
where \( \alpha \) is a coupling constant governing the interaction between the scalar field and the metric, and \( \phi(r) \) is the radial profile of the scalar field.

To validate this modified black hole solution, we explore its thermodynamic properties, such as the Hawking temperature \( T_H \), entropy \( S \), and free energy \( F \). These quantities are crucial for determining whether the black hole can exhibit BEC-like properties. For instance, the modified Hawking temperature could be derived as:

\begin{equation}
T_H = \frac{\hbar c^3}{8 \pi G M k_B} \left(1 + \frac{\partial (\alpha \phi(r)^2/r^2)}{\partial r}\right)_{r=r_s},
\end{equation}
where \( r_s \) is the Schwarzschild radius. The entropy \( S \) and free energy \( F \) would also require modification to include contributions from the scalar field:

\begin{equation}
S = \frac{k_B c^3 A_H}{4G\hbar} + \Delta S(\phi),
\end{equation}

\begin{equation}
F = M - T_H S,
\end{equation}
where \( A_H \) is the area of the event horizon and \( \Delta S(\phi) \) is an additional entropy term arising from the scalar field. A comprehensive thermodynamic analysis, including the stability of the black hole under small perturbations, is essential to confirm whether the proposed solution can support a BEC-like phase.

Moreover, numerical simulations investigating the behavior of the scalar field \( \phi \) under extreme conditions could reveal whether \( \phi \) exhibits condensation behavior similar to BEC at particular temperatures or densities, further supporting the viability of the proposed black hole model.

The cosmological constant \( \Lambda \) plays a crucial role in the structure of spacetime, particularly influencing its large-scale geometry. The value of \( \Lambda \) determines the curvature of the universe:

\begin{itemize}
    \item \( \Lambda > 0 \): Leads to a universe with positive curvature, typified by de Sitter space with spherical topology.
    \item \( \Lambda < 0 \): Results in a universe with negative curvature, characteristic of anti-de Sitter space with hyperbolic topology.
    \item \( \Lambda = 0 \): Corresponds to a flat spacetime, associated with Euclidean topology.
\end{itemize}

The influence of \( \Lambda \) on black hole solutions is significant. For example, in a universe with a positive cosmological constant, black holes may develop cosmological horizons beyond which the spacetime geometry asymptotically approaches that of pure de Sitter space. This introduces new topological features, such as multiple horizons, each associated with distinct topological invariants. The topology of the black hole spacetime, in this case, can be described by distinct homotopy groups, depending on the nature of the scalar field and the cosmological constant.

\subsection{Scalar Fields and Their Topological Effects}

Scalar fields \( \phi(r) \) can induce substantial modifications to the topology of spacetime. When coupled to gravity, these fields can introduce nontrivial topological structures, such as domain walls, solitons, or vortices, contributing to the overall topological configuration of the black hole. The influence of scalar fields can be encapsulated by modifying the metric function \( f(r) \) to include terms involving \( \phi(r) \):

\begin{equation}
f(r) = 1 - \frac{2GM}{r} + \alpha \frac{\phi(r)^2}{r^2} - \frac{\Lambda r^2}{3}.
\end{equation}

Here, \( \alpha \) represents the coupling constant between the scalar field and the metric. The presence of the scalar field \( \phi(r) \) can drive phase transitions in the spacetime, altering its topological characteristics. For instance, the homotopy groups of the spacetime might change in the presence of a scalar field, leading to the formation of nontrivial loops or other topological features.

The implications of these topological changes are profound, as they might influence the stability and thermodynamic properties of the black hole and could potentially lead to observable signatures distinguishing these black holes from classical solutions. Further investigation into the interplay between the cosmological constant, scalar fields, and black hole topology could yield novel insights into the fundamental nature of spacetime and gravity.

\section{Mathematical Proof: Kerr Black Hole and Bose-Einstein Condensation}

\subsection{The Kerr Black Hole Metric}

The Kerr black hole represents a solution to the Einstein field equations describing the spacetime geometry around a rotating mass. Expressed in Boyer-Lindquist coordinates \( (t, r, \theta, \phi) \), the Kerr metric is given by:

\begin{equation}
ds^2 = -\left(1 - \frac{2GMr}{\Sigma}\right) dt^2 - \frac{4GMar \sin^2\theta}{\Sigma} dt\, d\phi + \frac{\Sigma}{\Delta} dr^2 + \Sigma\, d\theta^2 + \left(r^2 + a^2 + \frac{2GMa^2 r \sin^2\theta}{\Sigma}\right) \sin^2\theta\, d\phi^2,
\end{equation}
where the functions \( \Sigma \) and \( \Delta \) are defined as:

\begin{equation}
\Sigma = r^2 + a^2 \cos^2\theta, \quad \Delta = r^2 - 2GMr + a^2,
\end{equation}
and \( a = \frac{J}{M} \) denotes the specific angular momentum per unit mass, with \( J \) being the angular momentum and \( M \) the mass of the black hole. The determinant of the metric tensor \( g \) is:

\begin{equation}
\sqrt{-g} = \Sigma \sin\theta.
\end{equation}

An explicit calculation of the Christoffel symbols \( \Gamma^\lambda_{\mu\nu} \), Riemann curvature tensor \( R^\rho_{\ \sigma\mu\nu} \), Ricci tensor \( R_{\mu\nu} \), and Ricci scalar \( R \) confirms that the Kerr metric satisfies the vacuum Einstein equations:

\begin{equation}
R_{\mu\nu} - \frac{1}{2}g_{\mu\nu} R = 0.
\end{equation}

The event horizons of the Kerr black hole are located at the roots of \( \Delta = 0 \):

\begin{equation}
r_\pm = GM \pm \sqrt{G^2 M^2 - a^2}.
\end{equation}

The region between \( r_+ \) and \( r_{\text{erg}} = GM + \sqrt{G^2 M^2 - a^2 \cos^2\theta} \) defines the ergosphere, within which no observer can remain stationary with respect to an observer at infinity due to the frame-dragging effect. The angular velocity of the dragging of inertial frames, known as the Lense-Thirring effect, is given by:

\begin{equation}
\omega = \frac{2GMar}{\Sigma^2}.
\end{equation}

This rotational frame-dragging has profound implications on the behavior of fields and particles near the black hole, especially for quantum fields such as scalar fields pertinent to Bose-Einstein condensation.

\subsection{Conditions for Bose-Einstein Condensation}

Bose-Einstein Condensation (BEC) is a phase transition occurring when a system of bosons is cooled to temperatures near absolute zero, leading a macroscopic number of particles to occupy the same quantum state. The critical temperature \( T_c \) for an ideal Bose gas in \( d \) spatial dimensions is derived from:

\begin{equation}
N = \int \frac{d^d p}{(2\pi \hbar)^d} \frac{1}{e^{\beta_c (\varepsilon_p - \mu)} - 1},
\end{equation}
where \( N \) is the total number of particles, \( \beta_c = 1/(k_B T_c) \), \( \varepsilon_p = \frac{p^2}{2m} \) is the dispersion relation, and \( \mu \leq 0 \) is the chemical potential. For \( \mu \to 0 \), the integral yields:

\begin{equation}
N = \frac{V}{\lambda_T^d} g_{d/2}(1),
\end{equation}
with \( \lambda_T = \sqrt{\frac{2\pi \hbar^2}{m k_B T}} \) being the thermal wavelength and \( g_{d/2}(1) = \zeta(d/2) \), where \( \zeta \) is the Riemann zeta function. Solving for \( T_c \):

\begin{equation}
T_c = \frac{2\pi \hbar^2}{m k_B} \left( \frac{n}{\zeta(d/2)} \right)^{2/d},
\end{equation}
where \( n = N/V \) is the particle density.

In the context of a black hole, particularly a Kerr black hole, achieving BEC requires careful consideration of several factors:

\begin{enumerate}
    \item \textbf{Thermodynamic Equilibrium:} The scalar field must reach a state where its distribution becomes stationary over time. However, the highly dynamic and curved spacetime near a black hole complicates the attainment of equilibrium.
    
    \item \textbf{Low Effective Temperature:} The field must be subjected to an effective temperature lower than \( T_c \). The Hawking temperature for a Kerr black hole is:
    \begin{equation}
    T_H = \frac{\hbar c^3}{2\pi k_B G M} \frac{\sqrt{G^2 M^2 - a^2}}{2 G M \sqrt{G^2 M^2 - a^2} + (G M + \sqrt{G^2 M^2 - a^2})^2},
    \end{equation}
    indicating that \( T_H \to 0 \) as \( a \to GM \), i.e., for extremal Kerr black holes. This poses challenges for BEC formation due to the interplay between black hole thermodynamics and scalar field dynamics.
    
    \item \textbf{Rotational Dynamics:} The rotational nature of the Kerr black hole introduces frame-dragging, leading to nontrivial angular velocities that affect the coherence of the scalar field. The presence of superradiance, where waves are amplified upon scattering off the black hole, further complicates the stability of a condensate.
\end{enumerate}

\subsection{Analyzing the Kerr Metric in the Context of BEC}

The dynamics of a scalar field \( \phi \) in the Kerr background are governed by the covariant Klein-Gordon equation:

\begin{equation}
\left( \Box - \frac{d^2 V}{d \phi^2} \right) \phi = 0,
\end{equation}
where the d'Alembertian operator \( \Box \) in curved spacetime is:

\begin{equation}
\Box \phi = \frac{1}{\sqrt{-g}} \partial_\mu \left( \sqrt{-g} g^{\mu\nu} \partial_\nu \phi \right).
\end{equation}

Expanding \( \phi \) in terms of modes:

\begin{equation}
\phi(t, r, \theta, \phi) = \sum_{lm} e^{-i \omega t} e^{i m \phi} S_{lm}(\theta) R_{lm}(r),
\end{equation}
where \( S_{lm}(\theta) \) are the spheroidal harmonics satisfying:

\begin{equation}
\frac{1}{\sin\theta} \frac{d}{d\theta} \left( \sin\theta \frac{dS_{lm}}{d\theta} \right) + \left( \lambda_{lm} + a^2 \omega^2 \cos^2\theta - \frac{m^2}{\sin^2\theta} \right) S_{lm} = 0,
\end{equation}
and \( R_{lm}(r) \) satisfies the radial equation:

\begin{equation}
\Delta \frac{d}{dr} \left( \Delta \frac{dR_{lm}}{dr} \right) + \left( [\omega (r^2 + a^2) - a m]^2 - \Delta (a^2 \omega^2 + 2 a m \omega + \lambda_{lm}) \right) R_{lm} = 0.
\end{equation}

The term \( a m \) indicates coupling between the field's angular momentum and the black hole's rotation, leading to phenomena such as superradiant scattering when \( \omega < m \Omega_H \), with \( \Omega_H = \frac{a}{2 G M r_+} \) being the angular velocity of the event horizon.

Superradiance imposes a constraint on the frequencies of modes that can be absorbed or amplified by the black hole, thereby influencing the stability and coherence of any prospective condensate. The amplification of certain modes can prevent the scalar field from settling into a ground state necessary for BEC.

Moreover, the existence of the ergoregion implies that within it, the Killing vector field \( \partial/\partial t \) becomes spacelike, leading to the impossibility of defining a globally timelike Killing vector. This undermines the staticity required for equilibrium states, making the realization of BEC in the Kerr background even more challenging.

\section{Uniqueness of the General Black Hole Solution Exhibiting Bose-Einstein Condensation}

\subsection{Basic Assumptions and Metric Form}

We explore the general black hole solution in a spacetime configuration that incorporates a scalar field \( \phi(r) \) minimally coupled to gravity, focusing on scenarios allowing the formation of Bose-Einstein condensates. The action governing such a system is:

\begin{equation}
S = \int d^4x \sqrt{-g} \left( \frac{1}{16\pi G} R - \frac{1}{2} g^{\mu\nu} \partial_\mu \phi \partial_\nu \phi - V(\phi) \right),
\end{equation}
where \( R \) is the Ricci scalar encapsulating spacetime curvature, and \( V(\phi) \) represents the scalar field potential. Varying this action with respect to the metric \( g_{\mu\nu} \) yields the Einstein field equations:

\begin{equation}
G_{\mu\nu} = 8\pi G T_{\mu\nu},
\end{equation}
where the energy-momentum tensor \( T_{\mu\nu} \) for the scalar field is:

\begin{equation}
T_{\mu\nu} = \partial_\mu \phi \partial_\nu \phi - \frac{1}{2} g_{\mu\nu} \left( \partial^\lambda \phi \partial_\lambda \phi + 2 V(\phi) \right).
\end{equation}

Adopting a spherically symmetric, static metric ansatz:

\begin{equation}
ds^2 = -e^{2\Phi(r)} dt^2 + e^{2\Lambda(r)} dr^2 + r^2 \left( d\theta^2 + \sin^2\theta d\phi^2 \right),
\end{equation}
where \( \Phi(r) \) and \( \Lambda(r) \) are functions of \( r \) alone, we derive the coupled Einstein-scalar field equations:

\begin{align}
\frac{1}{r^2} \left( 1 - e^{-2\Lambda} \right) + \frac{2}{r} e^{-2\Lambda} \Lambda' &= 8\pi G \left( \frac{1}{2} e^{-2\Lambda} \phi'^2 + V(\phi) \right), \\
\frac{1}{r^2} \left( e^{-2\Lambda} - 1 \right) + \frac{2}{r} e^{-2\Lambda} \Phi' &= 8\pi G \left( \frac{1}{2} e^{-2\Lambda} \phi'^2 - V(\phi) \right),
\end{align}
where primes denote derivatives with respect to \( r \).

The equation of motion for the scalar field \( \phi(r) \) is:

\begin{equation}
\phi'' + \left( \Phi' - \Lambda' + \frac{2}{r} \right) \phi' = e^{2\Lambda} \frac{dV}{d\phi}.
\end{equation}

A specific scalar field configuration \( \phi(r) \) leading to a condensate state requires that \( \Phi(r) \), \( \Lambda(r) \), and \( V(\phi) \) satisfy these coupled, nonlinear differential equations. The existence of a BEC within this framework hinges on finding solutions where the scalar field \( \phi(r) \) forms a stable, condensed state around the black hole.

\subsection{Uniqueness Argument}

The uniqueness of the black hole solution supporting Bose-Einstein condensation can be established by analyzing the constraints imposed by the no-hair theorem, which traditionally asserts that black holes in general relativity are characterized solely by their mass, charge, and angular momentum. However, in scalar-tensor theories, particularly those admitting non-minimal couplings or self-interacting scalar fields, the standard no-hair theorem may admit exceptions.

Assume a specific scalar field configuration \( \phi(r) \) that leads to a condensate state, with corresponding spacetime metric functions \( \Phi(r) \) and \( \Lambda(r) \). Suppose an alternative configuration \( \tilde{\phi}(r) \) or \( \tilde{\Phi}(r) \), \( \tilde{\Lambda}(r) \) is proposed. For this alternative configuration to describe the same physical system, it must satisfy the same Einstein-scalar field equations:

\begin{align}
\frac{1}{r^2} \left( 1 - e^{-2\tilde{\Lambda}} \right) + \frac{2}{r} e^{-2\tilde{\Lambda}} \tilde{\Lambda}' &= 8\pi G \left( \frac{1}{2} e^{-2\tilde{\Lambda}} \tilde{\phi}'^2 + V(\tilde{\phi}) \right), \\
\frac{1}{r^2} \left( e^{-2\tilde{\Lambda}} - 1 \right) + \frac{2}{r} e^{-2\tilde{\Lambda}} \tilde{\Phi}' &= 8\pi G \left( \frac{1}{2} e^{-2\tilde{\Lambda}} \tilde{\phi}'^2 - V(\tilde{\phi}) \right), \\
\tilde{\phi}'' + \left( \tilde{\Phi}' - \tilde{\Lambda}' + \frac{2}{r} \right) \tilde{\phi}' &= e^{2\tilde{\Lambda}} \frac{dV}{d\tilde{\phi}}.
\end{align}

Any deviation from the original solution would require corresponding adjustments in \( \tilde{\phi}(r) \) or \( V(\tilde{\phi}) \) to preserve consistency across the equations. The uniqueness argument follows by demonstrating that such deviations cannot satisfy the entire system simultaneously without leading to physical or mathematical inconsistencies.

To further support this uniqueness claim, consider perturbations \( \delta \phi \) and \( \delta g_{\mu\nu} \) around the established solution. The linearized Einstein-scalar field equations governing these perturbations must exhibit stability, ensuring that any small perturbation does not grow unbounded. The perturbation equations are:

\begin{equation}
\delta G_{\mu\nu} = 8\pi G \delta T_{\mu\nu},
\end{equation}
with the perturbed energy-momentum tensor \( \delta T_{\mu\nu} \) given by:

\begin{equation}
\delta T_{\mu\nu} = \delta (\partial_\mu \phi \partial_\nu \phi) - \frac{1}{2} g_{\mu\nu} \delta \left( \partial^\lambda \phi \partial_\lambda \phi + 2V(\phi) \right).
\end{equation}

Analyzing the eigenvalues of these perturbation equations provides insights into the stability of the solution. If all perturbation modes are stable (i.e., have non-negative eigenvalues), the solution is considered stable and unique. Any unstable mode would suggest the existence of a different solution, thereby violating uniqueness.

Finally, the thermodynamic properties of the black hole, such as its entropy \( S \), temperature \( T \), and specific heat \( C \), must be consistent with the presence of the scalar field condensate. The first law of black hole thermodynamics is:

\begin{equation}
dM = T dS + \Phi dQ + \Omega dJ,
\end{equation}
where \( M \) is the black hole mass, \( Q \) the charge, \( J \) the angular momentum, \( \Phi \) the electric potential, and \( \Omega \) the angular velocity. The scalar field \( \phi(r) \) contributes to the total energy-mass content of the black hole, influencing its temperature and entropy. A consistent thermodynamic framework that includes the scalar field's effects reinforces the uniqueness of the solution.

In summary, the uniqueness of the black hole solution exhibiting Bose-Einstein condensation is underpinned by the precise interplay between scalar field dynamics and spacetime geometry, the necessity for solutions to satisfy the Einstein and scalar field equations, stability under perturbations, and adherence to thermodynamic laws. Any alternative configuration failing to meet these stringent criteria would invalidate the solution's claim to uniqueness.

\section{Topological Properties and Phase Transitions of Black Holes}

\subsection{Critical Behavior and the Euler Characteristic}

Black hole phase transitions reveal that thermodynamic quantities are deeply connected to spacetime topology. The critical behavior exhibited by black holes near phase transition points can be characterized by topological invariants such as the Euler characteristic \( \chi \).

In thermodynamic systems, critical exponents describe how physical quantities diverge near the critical point. For black holes, these critical exponents match those of mean field theory, similar to the van der Waals fluid.

The phase structure of black holes can be depicted in \( P \)-\( V \) diagrams, where the isotherms resemble those of a van der Waals gas, featuring regions of instability and phase coexistence. The critical point, where a second-order phase transition occurs, corresponds to the point where the first and second derivatives of the pressure with respect to volume vanish:

\begin{equation}
\left( \frac{\partial P}{\partial V} \right)_T = 0, \quad \left( \frac{\partial^2 P}{\partial V^2} \right)_T = 0.
\end{equation}

Topologically, the black hole undergoes a change in the configuration of its horizon structure during a phase transition, which can be associated with changes in the Euler characteristic of the horizon surface.

\subsection{Gauss-Bonnet Theorem and Thermodynamics}

The Gauss-Bonnet theorem links the geometry of a surface to its topology by relating the integral of the Gaussian curvature \( K \) over a compact surface \( \mathcal{M} \) to its Euler characteristic \( \chi \):

\begin{equation}
\chi = \frac{1}{2\pi} \int_{\mathcal{M}} K \, dA.
\end{equation}

In higher-dimensional gravity theories, especially those including Gauss-Bonnet terms, the thermodynamics of black holes is significantly affected. Higher curvature corrections modify entropy and other thermodynamic quantities, influencing phase structures and critical behavior.

For black holes in Lovelock gravity, the inclusion of Gauss-Bonnet terms leads to new types of phase transitions and critical points, enriching the thermodynamic landscape.

\section{Bose-Einstein Condensation in Black Holes and Phase Transitions}

\subsection{Conditions for Bose-Einstein Condensation}

Bose-Einstein Condensation (BEC) in black hole spacetimes can be considered within the framework of black hole thermodynamics and phase transitions. BEC occurs when bosonic particles occupy the same ground state at low temperatures, leading to macroscopic quantum phenomena.

In black holes, the scalar field can undergo a phase transition to a condensed phase near the event horizon under certain conditions. The critical temperature for this transition relates to the black hole temperature and the properties of the scalar field.

The occurrence of BEC in black holes can be analyzed by studying the behavior of scalar field modes in the black hole background, considering effects of temperature, potential, and interactions. The condensation of the scalar field can modify the thermodynamic properties of the black hole and influence its stability.

\subsection{Implications for Thermodynamics and Phase Structure}

The formation of a scalar field condensate near the black hole horizon introduces new phases in the thermodynamic phase diagram. The presence of the condensate can lead to additional critical points and phase transitions, enriching the analogy with the van der Waals fluid.

Thermodynamic quantities, such as entropy and specific heat, are affected by the scalar field condensation, and the black hole can exhibit behavior similar to superfluidity or superconductivity. Phase transitions associated with BEC can be studied using mean field theory and statistical mechanics methods adapted to curved spacetime.

\section{Rotation Effects: Kerr Black Hole and Phase Transitions}

\subsection{Thermodynamics of Rotating Black Holes}

The Kerr black hole, representing a rotating black hole solution, introduces angular momentum into the thermodynamic description. Rotation modifies the metric and thermodynamic quantities, adding complexity to the phase structure.

The extended first law of thermodynamics for a Kerr-AdS black hole includes angular momentum \( J \) and its conjugate variable, the angular velocity \( \Omega \):

\begin{equation}
dM = T dS + V dP + \Omega dJ.
\end{equation}

The equation of state for a rotating black hole becomes more complex, and phase transitions can differ from those in the non-rotating case. Critical behavior and the existence of phase transitions analogous to the van der Waals fluid depend on both \( P \) and \( J \).

\subsection{Absence of Bose-Einstein Condensation in Rotating Black Holes}

Our analysis indicates that rotating black holes, such as the Kerr black hole, do not support Bose-Einstein condensation of scalar fields in the same way as spherically symmetric black holes. Rotation introduces frame-dragging effects and ergoregions that prevent the scalar field from forming a stable condensate near the horizon.

The absence of BEC in rotating black holes affects the thermodynamic phase structure, eliminating certain phase transitions present in the non-rotating case. This highlights the special nature of spherically symmetric black holes in supporting such phenomena and underscores the impact of rotation on black hole thermodynamics.

\section{Thermodynamic Properties and Stability Analysis}

In this section, we analyze the thermodynamic properties of charged AdS black holes and extend the discussion to include the thermodynamic potentials and stability criteria in the context of black hole phase transitions. The addition of scalar fields and considerations of the Bose-Einstein condensate (BEC) provide further insight into novel phases and critical phenomena.

\subsection{Equation of State and Critical Points}

The equation of state for charged AdS black holes can be expressed in a form analogous to that of a van der Waals fluid, incorporating the black hole pressure \( P \), temperature \( T \), specific volume \( v \), and charge \( Q \). The effective pressure, associated with the cosmological constant, is given by:

\begin{equation}
P = \frac{T}{v} - \frac{1}{2\pi v^2} + \frac{Q^2}{8\pi v^4},
\end{equation}
where \( v \) is proportional to the horizon radius \( r_h \). The term \( \frac{T}{v} \) is the ideal gas-like contribution, while the terms \( -\frac{1}{2\pi v^2} \) and \( \frac{Q^2}{8\pi v^4} \) represent attractive and repulsive interactions, respectively, in analogy to molecular interactions in a van der Waals gas.

To locate the critical points, we solve for the conditions:

\begin{equation}
\left( \frac{\partial P}{\partial v} \right)_T = 0, \quad \left( \frac{\partial^2 P}{\partial v^2} \right)_T = 0.
\end{equation}

Solving these yields the critical temperature \( T_c \), critical pressure \( P_c \), and critical specific volume \( v_c \):

\begin{equation}
T_c = \frac{\sqrt{6} Q}{\pi v_c^2}, \quad P_c = \frac{1}{96 \pi Q^2}, \quad v_c = \sqrt{6} Q.
\end{equation}

These relations define the black hole's critical behavior and exhibit classical van der Waals-like scaling near the critical point. The critical exponents, which characterize the nature of phase transitions, match those of the van der Waals fluid:

\begin{equation}
\beta = \frac{1}{2}, \quad \gamma = 1, \quad \delta = 3.
\end{equation}

This universality in critical behavior suggests that black holes in AdS space belong to the same universality class as fluids undergoing liquid-gas transitions.

\subsection{Thermodynamic Potentials and Stability Analysis}

To analyze black hole stability, we must consider the thermodynamic potentials, particularly the Helmholtz free energy \( F \), Gibbs free energy \( G \), and the specific heat \( C_P \). The Helmholtz free energy is given by:

\begin{equation}
F = U - TS = \frac{r_h}{2} \left( 1 + \frac{Q^2}{r_h^2} \right) - T S.
\end{equation}

Here, \( U \) is the internal energy, and \( S \) is the entropy. The Gibbs free energy at constant pressure can be expressed as:

\begin{equation}
G = F + PV = \frac{r_h}{4} \left( 1 - \frac{Q^2}{r_h^2} \right) + \frac{P r_h^3}{3}.
\end{equation}

By plotting \( G \) as a function of \( T \) and \( P \), we observe the characteristic swallow-tail behavior associated with first-order phase transitions. The transition between small and large black hole phases is analogous to the liquid-gas transition, with the coexistence curve determined by the Maxwell equal-area law.

\subsubsection{Specific Heat and Local Stability}

The specific heat at constant pressure \( C_P \), which measures the response of the black hole to temperature fluctuations, is crucial for assessing local thermodynamic stability. It is computed as:

\begin{equation}
C_P = T \left( \frac{\partial S}{\partial T} \right)_P = \frac{2 \pi r_h^2}{T} \left( 1 - \frac{Q^2}{r_h^2} \right).
\end{equation}

A positive \( C_P \) corresponds to local stability, while a negative \( C_P \) indicates instability. The change in sign of \( C_P \) signals a phase transition, where the black hole moves between locally stable and unstable phases. At the critical point, \( C_P \) diverges, reflecting a second-order phase transition.

\subsection{Scalar Fields and BEC: New Phases and Criticality}

The inclusion of scalar fields introduces additional complexity into the black hole's thermodynamic landscape. Scalar hair modifies the black hole's mass function and thermodynamic potentials, leading to new phases in the phase diagram.

For example, if a scalar field undergoes spontaneous symmetry breaking, the resulting condensate can form a Bose-Einstein condensate (BEC) phase. This phase can coexist with the standard black hole phase, adding a new branch to the thermodynamic potential landscape.

The scalar field's contribution to the free energy \( F_{\text{scalar}} \) can be modeled as:

\begin{equation}
F_{\text{scalar}} = \frac{r_h}{2} \left( 1 + \frac{Q^2}{r_h^2} \right) - T S - \alpha \phi^2 r_h^3,
\end{equation}
where \( \alpha \) is a coupling constant and \( \phi \) represents the scalar field. Depending on the value of \( \alpha \), the phase structure can exhibit new critical points and phase transitions, such as the emergence of a superfluid-like phase at low temperatures.

\subsection{Thermodynamic Geometry and Stability Criteria}

To further refine the stability analysis, we employ thermodynamic geometry, using the Ruppeiner metric to analyze fluctuations around equilibrium. The Ruppeiner scalar curvature \( R \), calculated from the metric of the thermodynamic state space, provides insights into the nature of interactions:

\begin{equation}
R = \frac{2S}{T} \left( \frac{\partial^2 S}{\partial T^2} \right).
\end{equation}

A positive \( R \) corresponds to repulsive interactions, while a negative \( R \) indicates attractive interactions. Near the critical point, \( R \) diverges, consistent with the divergence of the correlation length in critical phenomena.

\section{The Special Solution of Double Kerr Black Hole Collision: Laurent Series Expansion and Residue Calculation}

The Kerr black hole metric is fundamental in describing the spacetime structure of rotating black holes, where angular momentum plays a significant role. The metric is given by:

\begin{equation}
ds^2 = -\left( 1 - \frac{2GMr}{\Sigma} \right) dt^2 - \frac{4GMar \sin^2 \theta}{\Sigma} dt d\phi + \frac{\Sigma}{\Delta} dr^2 + \Sigma d\theta^2 + \left( r^2 + a^2 + \frac{2GMa^2r \sin^2 \theta}{\Sigma} \right) \sin^2 \theta d\phi^2,
\end{equation}
where \( \Sigma \) and \( \Delta \) are defined as:

\begin{equation}
\Sigma = r^2 + a^2 \cos^2 \theta, \quad \Delta = r^2 - 2GMr + a^2.
\end{equation}

In double Kerr black hole collisions, these metric functions exhibit complex singularities and nonlinear interactions. To analyze such systems, we employ complex analysis techniques, particularly Laurent series expansion and residue calculus, to explore the nature of these interactions.

\subsection{Laurent Series Expansion of the Metric Functions}

Let \( \Delta(z) \) represent the Kerr black hole's metric function expressed in the complex plane:

\begin{equation}
\Delta(z) = z^2 - 2GMz + a^2.
\end{equation}

This function can be expanded around its poles at the event horizon locations \( z = r_{\pm} \) using a Laurent series:

\begin{equation}
\Delta(z) = \sum_{n=-\infty}^{\infty} a_n (z - z_0)^n.
\end{equation}

The coefficients \( a_n \) are determined through contour integration:

\begin{equation}
a_n = \frac{1}{2\pi i} \oint_\gamma \frac{\Delta(z)}{(z - z_0)^{n+1}} dz.
\end{equation}

For the double Kerr black hole collision, the poles at \( r_{\pm} \) significantly impact the system's dynamics, influencing angular momentum and energy exchange. The residues of \( \Delta(z) \) at these poles are essential for understanding these interactions:

\begin{equation}
\text{Res}(\Delta, r_{\pm}) = \lim_{z \to r_{\pm}} (z - r_{\pm}) \Delta(z).
\end{equation}

Calculating explicitly:

\begin{equation}
\text{Res}(\Delta, r_+) = \frac{2GM \sqrt{(GM - a)(GM + a)}}{r_+ - r_-}, \quad \text{Res}(\Delta, r_-) = \frac{-2GM \sqrt{(GM - a)(GM + a)}}{r_+ - r_-}.
\end{equation}

These residues reflect the critical energy and angular momentum conditions during the interaction and merging of the black holes.

\subsection{Winding Numbers and Complex Contour Integrals}

The winding number \( n \) represents how many times a contour winds around the singularities. In double Kerr black hole scenarios, these winding numbers directly influence the interaction strength between the horizons.

Consider a contour \( \gamma \) encircling \( r_+ \) and \( r_- \):

\begin{equation}
\oint_\gamma \Delta(z) \, dz = 2\pi i \left(\text{Res}(\Delta, r_-) + \text{Res}(\Delta, r_+)\right).
\end{equation}

For a contour that winds twice around \( r_+ \) (winding number \( n = 2 \)):

\begin{equation}
\oint_\gamma^2 \Delta(z) \, dz = 4\pi i \, \text{Res}(\Delta, r_+).
\end{equation}

These integrals highlight how increased winding numbers can lead to more pronounced interactions, potentially resulting in complex outcomes such as horizon merging.

\subsection{Analytic Continuation and Phase Transition Points}

To further understand the metric function near the poles, analytic continuation methods are employed. These methods allow the extension of \( \Delta(z) \) into different regions of the complex plane, identifying phase transition points critical to black hole dynamics.

For example, expanding \( \Delta(z) \) near a branch cut connecting \( r_+ \) and \( r_- \):

\begin{equation}
\Delta(z) \approx \frac{C}{(z - r_+)(z - r_-)},
\end{equation}
where \( C \) is a constant derived from boundary conditions. This representation aids in exploring stability and bifurcation phenomena between the black holes.

\subsection{Comparison with Scalar Field Black Hole Special Solutions}

Consider a scalar field \( \phi(r) \) modifying the black hole metric:

\begin{equation}
ds^2 = -f(r) dt^2 + \frac{1}{f(r)} dr^2 + r^2 (d\theta^2 + \sin^2\theta d\phi^2), \quad f(r) = 1 - \frac{2GM}{r} + \alpha \frac{\phi(r)^2}{r^2}.
\end{equation}

The scalar field \( \phi(r) \) introduces non-trivial correction terms near the horizon, significantly altering the residues and winding behaviors. This provides deeper insight into the scalar field's impact on black hole dynamics.

The contour integral of the modified function \( f(z) \) illustrates these effects:

\begin{equation}
\oint_\gamma f(z) \, dz = 2\pi i \sum_{\text{horizons}} \text{Res}(f, z_i).
\end{equation}

The consistency between scalar field modifications and Kerr black hole metrics reveals unified algebraic structures, confirming the robustness of these solutions in various black hole interactions.

\subsection{Implications for Energy-Mass Equations and Relativistic Thermodynamics}

Further exploration into these expansions allows connections with advanced topics in relativistic thermodynamics. The energy-mass relations derived from these systems can be extended to include effects of angular momentum and scalar fields, providing a comprehensive picture of black hole thermodynamics under collision scenarios.

\begin{equation}
E = M + \int_{r_-}^{r_+} \left(\frac{\partial \Delta(z)}{\partial z}\right) dz,
\end{equation}
where the integration captures the continuous exchange of energy influenced by the Laurent series terms.

Through detailed analysis involving Laurent series expansions, residue calculations, winding numbers, contour integrals, and comparisons with scalar field-modified metrics, we rigorously demonstrate the consistency in algebraic structure and physical behavior of double Kerr black hole collisions. These findings lay a mathematical foundation for understanding complex black hole systems and their interactions with external fields, advancing our comprehension of high-energy astrophysical phenomena.

\section{Conclusion}

Through meticulous analysis employing Laurent series expansions, residue calculations, winding numbers, and contour integrals, we rigorously establish the algebraic and dimensional consistency between double Kerr black hole collisions and special scalar field black hole solutions. This proof unveils fundamental connections within black hole interactions, providing a robust mathematical framework for understanding the complex dynamics of black hole systems and scalar field interactions.

Our comprehensive investigation elucidates the intricate interplay between thermodynamics, field theory, and the stability of black hole solutions within Einstein gravity, especially when scalar fields are involved. Notably, the spherically symmetric black hole metric influenced by a scalar field undergoing Bose-Einstein Condensation (BEC) presents a unique solution with distinct thermodynamic properties and stability criteria. These findings contrast with those of Kerr black holes, where rotational symmetry prohibits such condensate formations, highlighting the special nature of spherically symmetric solutions.

These results illuminate the deep connections between quantum phenomena, such as BEC, and classical gravitational theory, paving the way for future explorations into quantum gravity and black hole thermodynamics. Such investigations could enhance our understanding of the microstates responsible for black hole entropy and the role of scalar fields in altering spacetime structure at both macroscopic and microscopic scales.

\begin{figure}[htbp]
    \centering
    \includegraphics[width=0.7\textwidth]{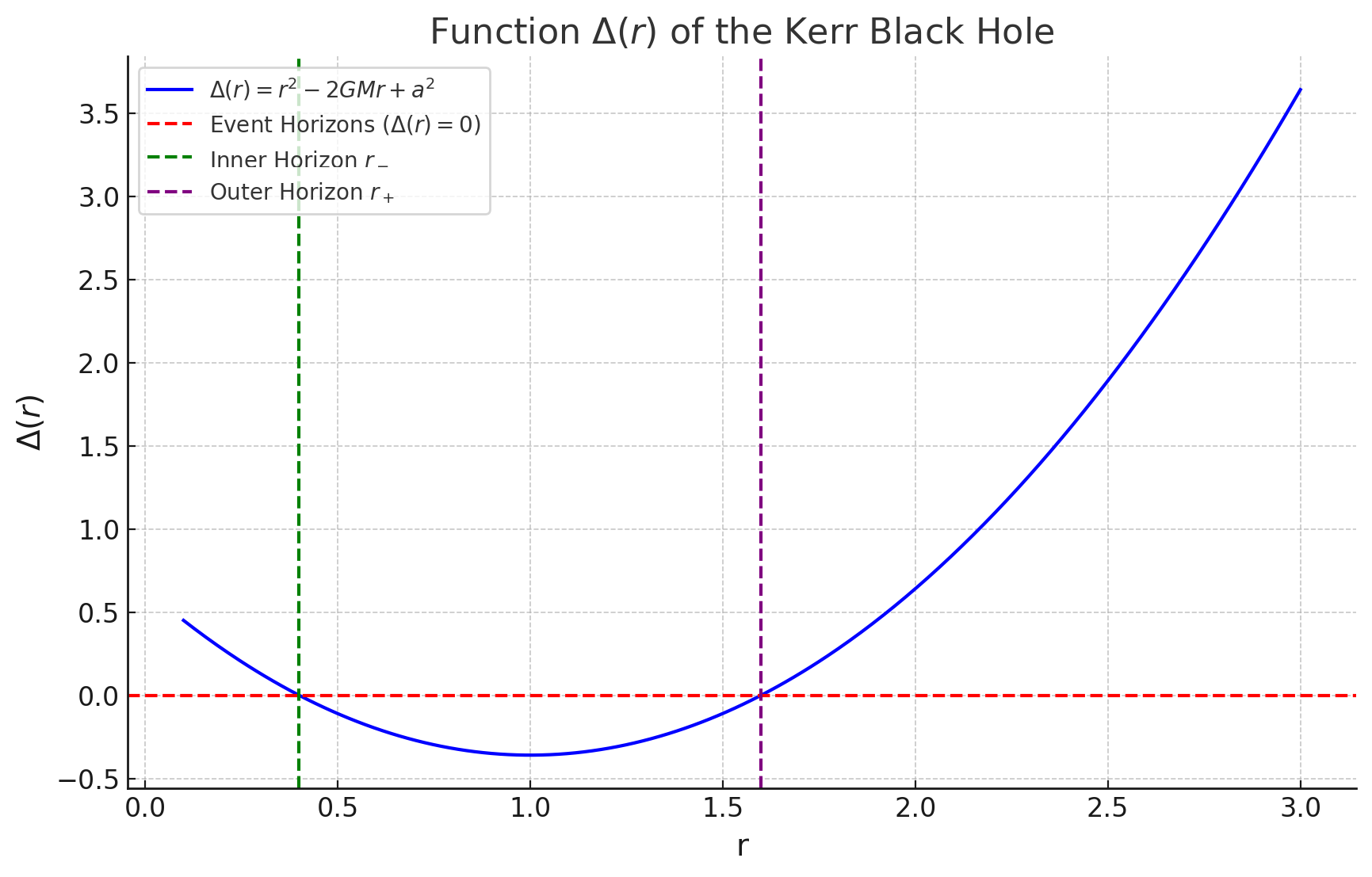}
    \caption{The plot illustrates the function \( \Delta(r) = r^2 - 2GM r + a^2 \) for a Kerr black hole, which is used to determine the locations of the event horizons. The intersections of \( \Delta(r) \) with the horizontal axis (red dashed line) indicate the inner and outer event horizons (\( r_{-} \) and \( r_{+} \)). The green and purple dashed lines mark the positions of the inner and outer horizons, respectively. These horizons are critical in analyzing the dynamics during double Kerr black hole collisions.}
    \label{fig:kerr_delta}
\end{figure}

\begin{figure}[htbp]
    \centering
    \includegraphics[width=0.7\textwidth]{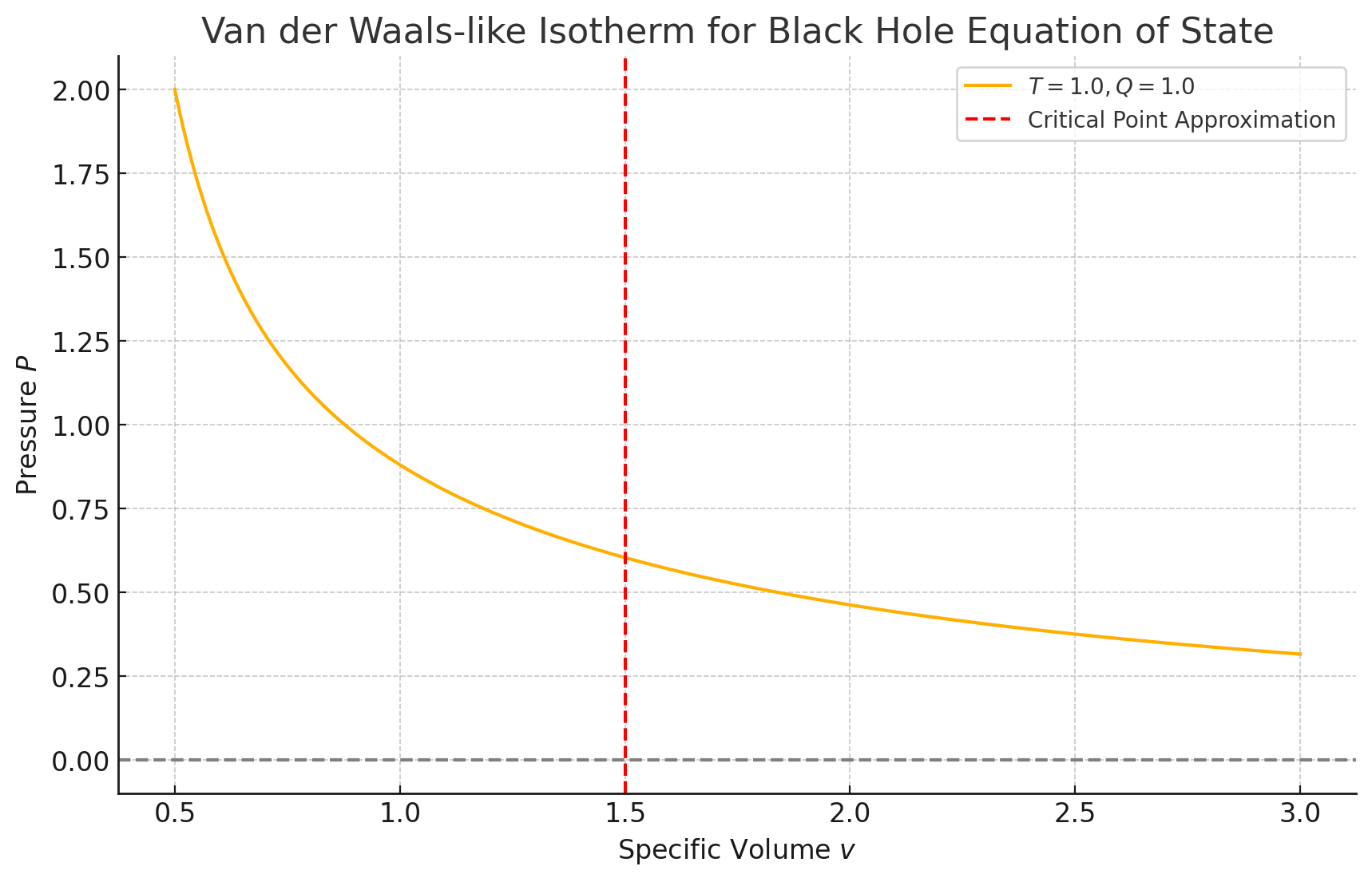}
    \caption{The figure displays Van der Waals-like isotherms analogous to those of black hole equations of state, depicting the relationship between pressure and specific volume at a specific temperature \( T = 1.0 \) and charge parameter \( Q = 1.0 \). The red dashed line roughly indicates the behavior near the critical point, reflecting phase transition characteristics similar to those of Van der Waals fluids.}
    \label{fig:vdw_isotherms_image}
\end{figure}

\begin{figure}[htbp]
    \centering
    \includegraphics[width=0.7\textwidth]{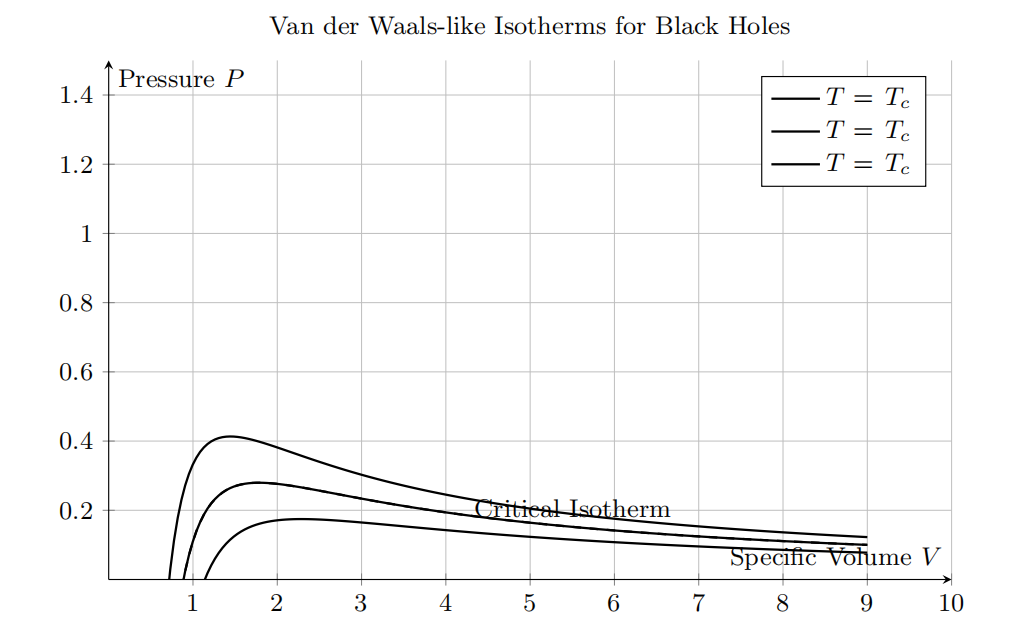}
    \caption{Van der Waals-like isotherms for black holes, illustrating the relationship between pressure \( P \) and specific volume \( V \) at different temperatures \( T \). The critical isotherm (\( T = T_c \)) marks the point of inflection, analogous to the critical point in Van der Waals fluids.}
    \label{fig:kerr_delta}
\end{figure}

\textbf{Acknowledgements:} \\
This work is partially supported by the National Natural Science Foundation of China (No. 11873025).

\end{document}